\documentclass[12pt]{iopart}
\usepackage{graphicx}
\usepackage{amssymb}
\usepackage{xcolor}
\newcommand{\bb}[1]{\mathbb{#1}}

\begin{document}
	
	\title[Aqua-PACMANN]{A New Paradigm of Reservoir Computing Exploiting Hydrodynamics}
	
	\author{Giulia Marcucci$^*$, Piergiorgio Caramazza, Shamit Shrivastava}
	\address{Apoha Ltd, 242 Acklam Rd, London W10 5JJ, UK}
	\ead{$^*$irreversiblegm@gmail.com}
	\vspace{10pt}
	\begin{indented}
		\item[] \today
	\end{indented}
	
	\begin{abstract}
		%Opening:
		Nonlinear waves have played a historical role in laying the foundations
		of the science of complexity. Recently, they have also allowed the development of a new reservoir computing paradigm: neuromorphic computing by waves. %~\cite{2020Marcucci}.
		%Challenge
		In these systems, the information transmission acts as the excitation of wave dynamics, whose evolution processes the information to perform complex tasks at low energy consumption.
		%Action:
		To enable nonlinear hydrodynamic waves to do computing, we designed the Aqua-Photonic-Advantaged Computing Machine by Artificial Neural Networks (Aqua-PACMANN), a system where wave propagation in shallow water is the leading physical phenomenon, and the presence of electronics can be reduced to a CCD camera in detection.
		%Resolution:
		We show the realization of an XNOR logic gate as proof of concept of the Aqua-PACMANN's architecture and pave the way to a new class of fluid dynamic neuromorphic computing.
	\end{abstract}
	
	%\maketitle

	%---------------------------
	\section{Introduction}
	\label{sec:intro}
	%---------------------------
	
	%\bred{\textit{Why this work?}}
	
	In the analog and unconventional computing community, neuromorphic computing~(NMC) has been gaining rapidly increasing attention. Its transformative purpose of engineering machines based on brain-like algorithms, such as artificial neural networks~(ANNs)~\cite{2011Langley}, can lead to a new industrial revolution where computers are able to solve problems following a holistic approach because their performance is directly embedded into the complexity of a physical system~\cite{2014Colombo}.
	
	The leading paradigm of NMC capable of bringing such innovation is reservoir computing~(RC)~\cite{2019Tanaka}. The last two decades have witnessed incredible efforts in developing RC in every possible direction. From a modeling perspective, the liquid state machines~(LSM)~\cite{2011Maass}, echo state network~(ESNs)~\cite{2004Jaeger}, and extreme learning machine~(ELM)~\cite{2006Huang} paved the way to the first attempts to realize a new generation of hardware, whose core is no more a silicon chip, but the propagation of waves in a physical reservoir~\cite{2003Fernando,2020Marcucci,2022Ma}.
	
	In a hydrodynamic system, a hardware implementation of the LSM was realized at the beginning of the millennium~\cite{2003Fernando}. This system uses mechanical excitation of linear waves in a bucket filled with water and, by imaging the wave interference at the center of the bucket, it accomplishes complex tasks such as vowel recognition or logic operations. However, it does not exploit any hydrodynamic nonlinearity to enhance the complexity degree of the reservoir and get a potential Turing-complete information processing~\cite{2021Cardona}.
	
	%\bred{\textit{What is this work?}}
	
	In this paper, we show how nonlinear waves in shallow water realize an ELM.
	It is well known that water waves are described by the Navier-Stokes equations~(NSE), and that their $(1+1)-$dimensional approximation in shallow water gives rise to the Korteweg-de Vries~(KdV) equation~\cite{1981Miles}. Another approximation of the NSE is the nonlinear Schr\"odinger equation, which can describe wave propagation in deep water~\cite{1983Peregrine}.
	It has already been shown that the nonlinear Schr\"odinger equation can model the reservoir of an ELM~\cite{2020Marcucci}, breaking new ground to design new RC devices governed by the NSE and capable of performing interpolation, classification, and logic gates. 
	
	Here, we demonstrate how superpositions of chosen KdV cnoidal solutions can encode information and their collision with a KdV soliton can process it. 
	In the next section, we describe our general Aqua-PACMANN architecture in terms of ESNs. In the case of a KdV, this system operates in its fully feedforward limit, given rise to an ELM. We show how the KdV theory is embedded in our network determining the information encoding and processing.
    The resulting design of a logic gate is illustrated in Sec.~\ref{sec:XNOR}. Here, plots of the three Aqua-PACMANN modules, namely, encoding, processing, and decoding, are reported, demonstrating the link between our ANN and KdV waveforms. 
	To conclude, Sec.~\ref{sec:exp} describes a possible physical realization of our system including numbers and dimensions.
    %and we trace the path to go beyond the ELM and design an ESN. Indeed, the KdV-Burgers-Kuramoto~(KBK) equation extends the KdV introducing losses in our system, which we can tune to achieve recurrence~\cite{2019Alimirzaluo}. 
	
	%---------------------------
	\section{Neuromorphic Architecture for a Logic Gate}
	\label{sec:NMC}
	%---------------------------
	
\begin{figure}[h!]
		\begin{center}
			\includegraphics[width=\linewidth]{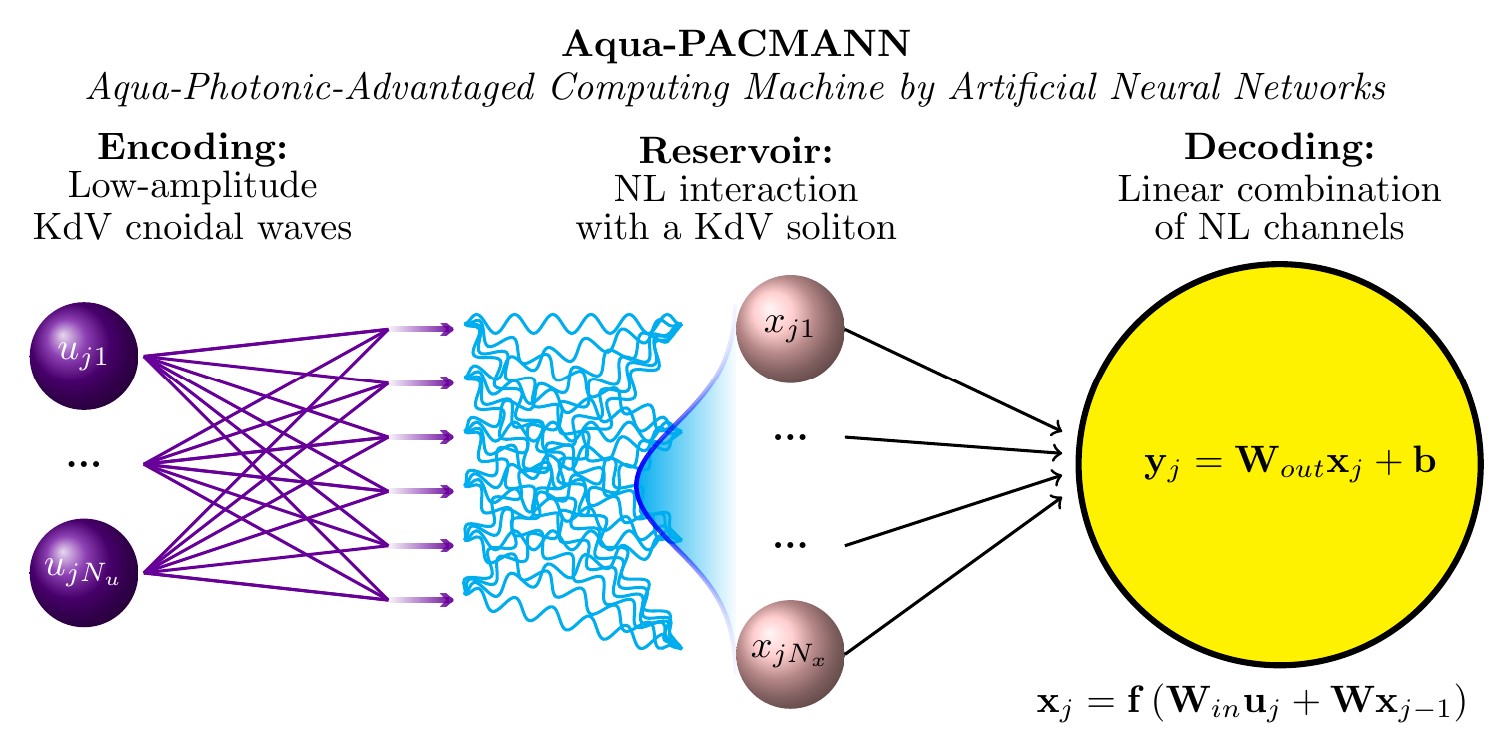}
			\caption{Aqua-PACMANN artificial neural network pictorial representation. The first layer is the encoding, composed of $N_u$ truncated low-amplitude cnoidal waves with wave numbers and amplitudes representing data labels and values, respectively. These waves propagate in shallow water and collide with a KdV soliton, constituting the reservoir. The reservoir readout is obtained by measuring the water height in a fixed detection point at $N_x$ different times. The decoding layer is a linear combination of the readout with trained weights, added to a bias vector.}
			\label{fig:network}
		\end{center}
	\end{figure}

	In the classical domain, the most relevant RC architectures are ESNs, LSM, and ELM. Their models can be generalized to
	\begin{eqnarray}
		\mathbf{y}(t_j) & = & \mathbf{W}_{out}\mathbf{x}(t_j) + \mathbf{b},\label{eq:RCy}\\
		\mathbf{x}(t_j) & = & f\left[\mathbf{W}_{in}\mathbf{u}(t_j) + \mathbf{W}\mathbf{x}(t_{j-1})\right],\label{eq:RCx}
	\end{eqnarray}  
	with $\mathbf{y}(t_j)$ the network output at time $t_j$ satisfying the condition
	\begin{equation}
		||\mathbf{y}(t_j)-\mathbf{y}_{T}(t_j)||<\epsilon \;\; \forall j=1,...,N,
		\label{eq:RC_training}
	\end{equation}
	where $\mathbf{y}_{T}$ is the target output and $\epsilon\gtrsim 0$.
	Figure~\ref{fig:network} provides a sketch of the implementation of such ANN in Aqua-PACMANN. Here, $\mathbf{u}(t_j)$ and $\mathbf{x}(t_j)$ are vectors of nodes in the input and reservoir readout layers at time $t_j$, respectively. Specifically, $u_k(t_j):=u_{jk},$ $k=1,...,N_u$ is a KdV low-amplitude cnoidal wave and $x_i(t_j):=x_{ji},$ $i=1,...,N_x$ is the detected height of water in a spatiotemporal point $x_D$. Furthermore, $\mathbf{W}_{out}$ is the matrix of trainable weights, $\mathbf{b}$ is the bias vector, $f$ is the activation function, and $\mathbf{W}_{in}$ and $\mathbf{W}$ are the matrices of the non-trainable weights of the reservoir.
	Eq.~(\ref{eq:RC_training}) allows us to obtain the weights matrix $\mathbf{W}_{out}$ in Eq.~(\ref{eq:RCy}) as 
	\begin{equation}
		\mathbf{W}_{out} = \left(\mathbf{y}_{T}(t_1)-\mathbf{b}\;\; \dots \;\; \mathbf{y}_{T}(t_N)-\mathbf{b}\right)\cdot\left(\mathbf{x}(t_1) \;\; \dots \;\; \mathbf{x}(t_N)\right)^{\dagger},
		\label{eq:Wout}
	\end{equation}
	where $\dagger$ stands for the Moore-Penrose pseudo-inverse matrix.
	
	Generically, the system defined by Eqs.~(\ref{eq:RCy},\ref{eq:RCx}) models a ESN. In the limit case $\mathbf{W}=\mathbf{0}$, one gets the non-recurrent, feedforward version of ESNs, namely, the ELM. On the other hand, when Eq.~(\ref{eq:RCx}) is such that $\mathbf{x}$ represents spike neurons, Eqs.~(\ref{eq:RCy},\ref{eq:RCx}) describe a LSM.
	
	To realize Aqua-PACMANN, we exploit the complexity of the KdV equation and its solutions interaction~\cite{2015Boyd}.
	When considering a water wave propagating along $x$ in time $t$, such that it is stationary along the transverse direction $y$, the wave normal velocity $u=u(x,t)$ (i.e., along $z$), which expresses its height instantaneous variations, follows the KdV equation
	\begin{equation}
		\partial_t u + u \partial_x u + \beta \partial_x^3 u = 0,
		\label{eq:kdv}
	\end{equation}
	with $\beta\in\bb{R}^+$ the dispersion coefficient, and $\partial_t:=\frac{\partial}{\partial t}$ the derivative operator.
	The traveling wave ansatz $u(x,t)=g\left[c_0(x-vt)\right]$, with $c_0$ the wave number and $v$ the speed, allows us to compute the KdV solutions known as cnoidal waves 
	\begin{equation}
		u = r_1-(r_1-r_2)\mbox{ sn}^2\left[\sqrt{\frac{r_1-r_3}{12\beta}}(x-vt),\,\frac{r_1-r_2}{r_1-r_3}\right].
		\label{eq:cnoidal}
	\end{equation}
	In Eq.~(\ref{eq:cnoidal}), $\mbox{sn}(\theta,m)$ is the Jacobi elliptic sine of argument $\theta$ and parameter $m\in\left[0,1\right]$; the parameters $r_1 \geq r_2 \geq r_3$ are the Riemann invariants associated to the KdV Riemann problem~\cite{2011Ablowitz} and are related to the wave speed as it is their arithmetic mean
	\begin{equation}
		v = \frac{r_1+r_2+r_3}{3}.
		\label{eq:vel}
	\end{equation}
	Without loss of generality, we also derive the normalized version of Eq.~(\ref{eq:kdv})
	\begin{equation}
		\partial_{\tau} \phi + \phi \partial_{\xi} \phi + \nu \partial_{\xi}^3 \phi = 0,
		\label{eq:nkdv}
	\end{equation}
	with only adimensional variables, obtained from the previous ones as ratios over references quantities such as
	\begin{equation}
		\tau := \frac{t}{T}, \; \xi := \frac{x}{v_0T}, \; \phi := \frac{u}{v_0}, \; \nu := \frac{\beta}{v_0^3T^2}.
		\label{eq:normalization}
	\end{equation}
	Hereafter, all the variables will be considered in arbitrary units~(a.u.), keeping in mind that Eq.~(\ref{eq:normalization}) enables us to move from adimensional quantities to whatever system of measurements.
	
	Traveling waves in Eq.~(\ref{eq:cnoidal}) have two limit regimes, depending on $0 \leq \frac{r_1-r_2}{r_1-r_3} \leq 1$:
	\begin{enumerate}
		\item $r_1 \gtrsim r_2 \gg r_3 \Longrightarrow \frac{r_1-r_2}{r_1-r_3} \sim 0$. In this case, solutions in Eq.~(\ref{eq:cnoidal}) can be approximated to the low-amplitude waves
		\begin{equation}
			u \simeq r_1-\epsilon\mbox{ sin}^2\left[\sqrt{\frac{r_1-r_3}{12\beta}}(x-vt)\right],
			\label{eq:lowampl}
		\end{equation}
		with $\epsilon := r_1-r_2$.
		\item $r_1 > r_2 = r_3 \Longrightarrow \frac{r_1-r_2}{r_1-r_3} = 1$. This limit defines the KdV soliton
		\begin{equation}
			u_s = r_2 + (r_1-r_2)\mbox{ sech}^2\left[\sqrt{\frac{r_1-r_2}{12\beta}}(x-vt)\right].
			\label{eq:soliton}
		\end{equation}
	\end{enumerate}
	We use trains of (i) to encode information in Aqua-PACMANN, and (ii) to build the nonlinear reservoir, as sketched in Fig.~\ref{fig:network}.
	
	Let us define the superposition of waves
	\begin{equation}
		u = \tilde{u}_e + u_s, \; \tilde{u}_e = \sum_{n=1}^{N_u} \tilde{u}_n,\;
		\tilde{u}_n := \epsilon_n\mbox{cos}^2\left[k_n(x-v_n t)\right]
		\label{eq:train}
	\end{equation}
	with $u_s$ defined in Eq.~(\ref{eq:soliton}), and $v_n = r_2 +\frac{2}{3}\epsilon_n-4\beta k_n^2$. It turns out that $u$ in Eq.~(\ref{eq:train}) satisfies Eq.~(\ref{eq:kdv}), thus $u(x, t=0)$ can be chosen to encode information in our bucket filled with shallow water. However, to take into account the finite excitation of water waves in a real experiment, we choose to use its smooth truncated version
	\begin{equation}
		u_0 (x)= {u}_e(x) + u_s(x+L,0), \; u_e = e^{-\left(\frac{2x}{l}\right)^8}\sum_{n=1}^{N_u} \bar{u}_n,\;
		\bar{u}_n := \epsilon_n\mbox{cos}^2\left(k_n x\right)
		\label{eq:truetrain}
	\end{equation}
	with $L$ expressing the delay $L/v$ between the encoding waves and the soliton, and $l$ the length of the encoding waves.

    In Sec.~\ref{sec:XNOR}, we show how Eqs.~(\ref{eq:kdv},\ref{eq:truetrain}) allow us to realize an XNOR gate. In general, we encode the $j$-th set of data in $u_j(x,t=0)$ defined as $u_0$ in Eq.~(\ref{eq:truetrain}). The information about the nature of the $n$-th element of the $j$-th set of data is encoded in the wave number of the $n$-th encoding wave, that is, $k_n$ is the label of the $n$-th item, and its value is encoded in the amplitude $\epsilon_n$. Then, we let $u_j$ propagate into shallow water and we detect the resulting water instantaneous height $u_j(x,t)$ in a point $x=x_D$ at times $t_1<t_2<...<t_{N_x}$, in order to build the readout vector
    \begin{equation}
        \mathbf{x}_j:=\left(\begin{array}{c}
        u_j(x=x_D, t=t_1)\\
        u_j(x=x_D, t=t_2)\\
        \vdots\\
        u_j(x=x_D, t=t_{N_x})
        \end{array}\right).
    \label{eq:readoutvec}
    \end{equation}
    Being $v > v_n\;\forall\, n=1,\dots,N_u$, the encoding waves collide with the soliton in a spatiotemporal point $\left(\frac{v_n}{v-v_n}L,\frac{L}{v-v_n}\right)$. To maximize the nonlinearity of our system, we choose the detection point $x_D$ in a surrounding of all the collision spatial coordinates $\frac{v_n}{v-v_n}L$.
    By repeating this measurement for all the data in the training set, we get the $N_x \times N$ response matrix
    \begin{equation}
        \mathbf{X}=\left(\begin{array}{ccc}
        u_1(x=x_D, t=t_1) & \dots & u_N(x=x_D, t=t_1)\\
        \vdots & \ddots & \vdots\\
        u_1(x=x_D, t=t_{N_x}) & \dots & u_N(x=x_D, t=t_{N_x})
    \end{array}\right).
    \label{eq:response}
    \end{equation}
    
    Our XNOR gate works in the exact learning conditions, which are $N_x=N$ and $\det\left(\mathbf{X}\right)\neq 0$. These conditions have specific geometric meaning~\cite{2006Huang}, and the latter is strictly related to the nonlinearity of the system~\cite{2020Marcucci}. This justifies the excitation of the KdV soliton in our bucket.
    From Eq.~(\ref{eq:Wout}), we get the definition of the weights matrix for the exact learning in the case $\mathbf{b}=\mathbf{0}$
    \begin{equation}
        \mathbf{W}_{out}=\mathbf{Y}_T\mathbf{X}^{-1},
        \label{eq:exactlearning}
    \end{equation}
	where $\mathbf{Y}_T$ is the target matrix.
 
	%---------------------------
	\section{XNOR Gate}
	\label{sec:XNOR}
	%---------------------------
	
	We use Aqua-PACMANN to realize a logic gate with truth table in Table~\ref{tab:xnor_truthtable}, that is, an XNOR.	
	\begin{table}[h!]
		\begin{center}
			\begin{tabular}{|c|c|c|}
				\hline
				$A$ & $B$ & $A \odot B$ \\
				\hline
				$1$ & $1$ & $0$ \\
				$1$ & $0$ & $1$ \\
				$0$ & $1$ & $1$ \\
				$0$ & $0$ & $0$ \\
				\hline
			\end{tabular}
			\caption{XNOR truth table.}
			\label{tab:xnor_truthtable}
		\end{center}
	\end{table}
	
\begin{figure}[h!]
		\begin{center}
			\includegraphics[width=\linewidth]{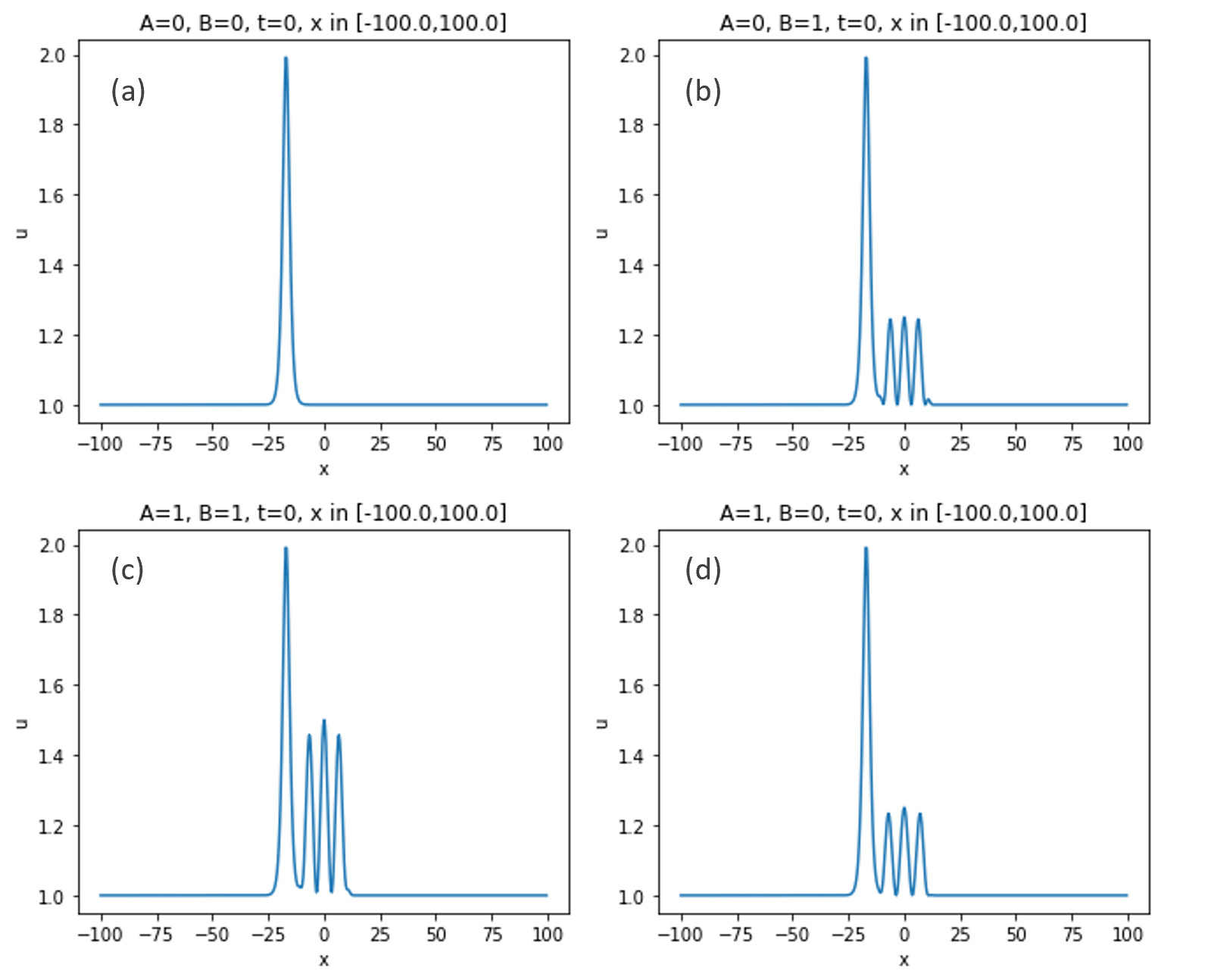}
			\caption{XNOR encoding waves (defining parameters in Table~\ref{tab:enc}) and soliton [defining parameters in Eq.~(\ref{eq:solparameters})] at $t=0$ in the four combinations defined in Table~\ref{tab:xnor_truthtable}. Titles in each panel point at the encoded Boolean variable couple.}
			\label{fig:enc}
		\end{center}
	\end{figure}

	Profiles of each combination of Eq.~(\ref{eq:truetrain}) with $N_u=2$ for the XNOR, with the whole set of encoding waves and the soliton, are reported in Fig.~\ref{fig:enc}. Here, the parameters in a.u. that define the soliton [Eq.~(\ref{eq:soliton})] at $t=0$ are
    \begin{equation}
        r_1=2,\;\;\; r_2=r_3=1,\;\;\; \beta=1/3,\;\;\; L=17,
        \label{eq:solparameters}
    \end{equation}
    with the resulting velocity $v=\frac{4}{3}$, while the parameters defining the encoding waves are $l=20$, $k_1=\frac{\sqrt{3}}{4}$ for the Boolean variable $A$, $k_2=\frac{1}{2}$ for the Boolean variable $B$, $\epsilon_1=\frac{1}{4}$ for the Boolean value 1 (i.e, true), $\epsilon_2=0$ for the Boolean value 0 (i.e., false), as summarized in Table~\ref{tab:enc}.
    \begin{table}[h!]
		\begin{center}
			\begin{tabular}{|c|c|c|c|}
				\hline
				$A$ & $B$ & $\bar{u}_1= \epsilon_1\mbox{cos}^2\left(k_1 x\right)$, $k_1=\frac{\sqrt{3}}{4}$ & $\bar{u}_2= \epsilon_2\mbox{cos}^2\left(k_2 x\right)$, $k_2=\frac{1}{2}$ \\
				\hline
				$1$ & $1$ & $\epsilon_1=\frac{1}{4}$, resulting speed $=\frac{11}{12}$ & $\epsilon_2=\frac{1}{4}$, resulting speed $=\frac{5}{6}$\\
				$1$ & $0$ & $\epsilon_1=\frac{1}{4}$, resulting speed $=\frac{11}{12}$ & $\epsilon_2=0$, absence of $\bar{u}_2$\\
				$0$ & $1$ & $\epsilon_1=0$, absence of $\bar{u}_1$ & $\epsilon_2=\frac{1}{4}$, resulting speed $=\frac{5}{6}$\\
				$0$ & $0$ & $\epsilon_1=0$, absence of $\bar{u}_1$ & $\epsilon_2=0$, absence of $\bar{u}_2$\\
				\hline
			\end{tabular}
			\caption{XNOR encoding waves defining parameters.}
			\label{tab:enc}
		\end{center}
	\end{table}

\begin{figure}[h!]
		\begin{center}
			\includegraphics[width=\linewidth]{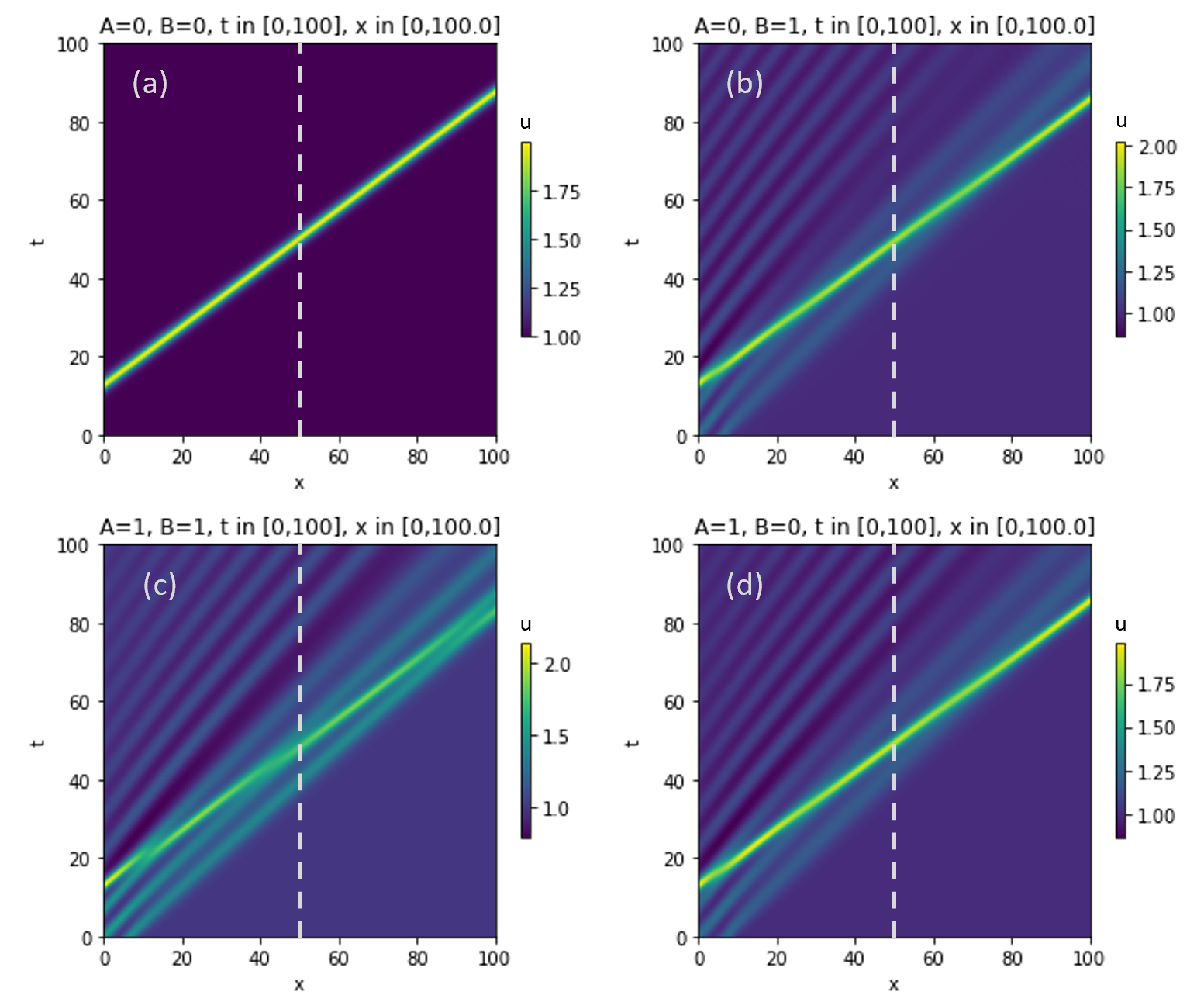}
			\caption{Information processing realized as numerical simulations of the KdV equation [Eq.~(\ref{eq:kdv})] with initial conditions shown in Fig.~\ref{fig:enc}. Titles in each panel point at the encoded Boolean variable couple. The four transverse profiles at $x=x_D=50$ along the white dashed lines are reported in Fig.~\ref{fig:dec}.}
			\label{fig:proc}
		\end{center}
	\end{figure}

    These four profiles represent the initial conditions of the related KdV Cauchy problem, namely, the Aqua-PACMANN's reservoir sketched in Fig.~\ref{fig:network}. Figure~\ref{fig:proc} shows the numerical simulations of their evolution. In a RC architecture, this is the stage in which the information gets processed. It has been previously shown that the information processing stage needs a non-trivial degree of nonlinearity to build an invertible response matrix $\mathbf{X}$ [Eq.~(\ref{eq:response})], that is, to enable exact learning~\cite{2020Marcucci}. The collisions of the encoding waves with the KdV soliton provide the right physical conditions to perform a network training as defined in Eq.~(\ref{eq:exactlearning}).
	
\begin{figure}[h!]
		\begin{center}
			\includegraphics[width=\linewidth]{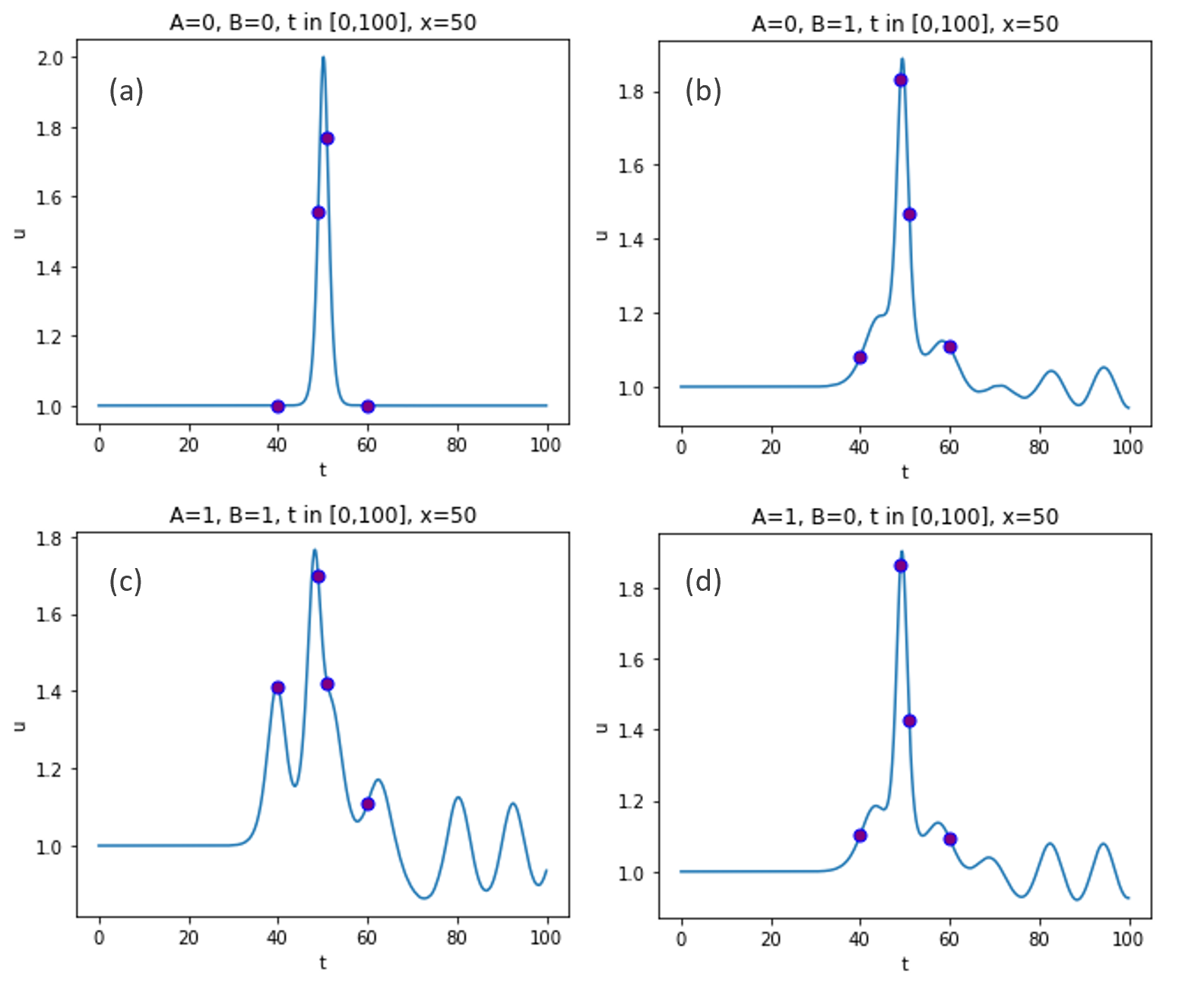}
			\caption{Transverse profiles at $x=x_D=50$, that is, along the white dashed lines in Fig.~\ref{fig:proc}, obtained by simulating the KdV equation [Eq.~(\ref{eq:kdv})] with initial conditions shown in Fig.~\ref{fig:enc} in the logic gate four cases (Table~\ref{tab:enc}). Each panel shows four purple dots at detection times $t_1=40, t_2=49, t_3=51, t_4=60$ representing the components of the readout vectors defined in Eq.~(\ref{eq:readoutvec}). Titles in each panel point at the encoded Boolean variable couple.}
			\label{fig:dec}
		\end{center}
	\end{figure}

	The last layer or the reservoir provides the readout vectors defined in Eq.~(\ref{eq:readoutvec}), with components obtained by measuring the water instantaneous height at detection point $x_D=50$ and times $t_1=40, t_2=49, t_3=51, t_4=60$. Figure~\ref{fig:dec} shows the resulting four transverse profiles at $x=x_D$ and measure points. As explained in Sec.~\ref{sec:NMC}, they allow us to obtain the response matrix
    \begin{equation}
		\mathbf{X} =
		\left(
		\begin{array}{cccc}
			1.0000 & 1.0798 & 1.1021 & 1.4108 \\
			1.5545 & 1.8292 & 1.8641 & 1.6995 \\
			1.7659 & 1.4670 & 1.4253 & 1.4211 \\
			1.0000 & 1.1078 & 1.0931 & 1.1087
		\end{array}
		\right),
		\label{eq:XNORresponse}
	\end{equation}
    with $\det\left(\mathbf{X}\right) = -0.0115\neq 0$. More specifically, vertical coordinates of purple dots in Fig.~\ref{fig:dec}a are the components of $\mathbf{X}$ first column, vertical coordinates of purple dots in Fig.~\ref{fig:dec}b are the components of $\mathbf{X}$ second column, and so on, Figs.~\ref{fig:dec}d,c correspond to the third and fourth column, respectively.

    To finalize the training, we define the target matrix $\mathbf{Y}_{T}$ on the canonical basis of $\mathbb{R}^2$ as
    	\begin{equation}
		\mathbf{Y}_{T} =
		\left(
		\begin{array}{cccc}
			1 & 0 & 0 & 1 \\
			0 & 1 & 1 & 0
		\end{array}
		\right).
		\label{eq:target}
	\end{equation}
	In this representation, $\left(\begin{array}{c}1\\0\end{array}\right)$ means \textit{false}, $\left(\begin{array}{c}0\\1\end{array}\right)$ means \textit{true}, and the order of $Y_T$ columns follows the third column in Table~\ref{tab:xnor_truthtable}. Eq.~(\ref{eq:exactlearning}) allows us to compute the XNOR weights matrix	
	\begin{equation}
		\mathbf{W}_{out} =
		\left(
		\begin{array}{cccc}
			2.8695 & -1.0731 & 1.8655 & -3.4956 \\
			-2.7375 & 1.3158 & -1.5995 & 3.5165
		\end{array}
		\right).
		\label{eq:weights}
	\end{equation}
	Now the training is complete and the Aqua-PACMANN XNOR is fully defined. The truth table~\ref{tab:xnor_truthtable} can now be obtained by encoding the Boolean variable $A,B$ in the system as previously explained, measuring the readout $\mathbf{x}_j$ at $x=x_D=50$ and $t_1=40, t_2=49, t_3=51, t_4=60$ and computing 	
	\begin{equation}
		\mathbf{y}_j = \mathbf{W}_{out}\mathbf{x}_j.
		\label{eq:linregr}
	\end{equation}
	The resulting $\mathbf{y}_j$ is $\mathbb{R}^2$ canonical versor such that $\left(\begin{array}{c}1\\0\end{array}\right)$ means \textit{false}, and $\left(\begin{array}{c}0\\1\end{array}\right)$ means \textit{true}, with an error of $\pm 10^{-3}$.

    %---------------------------
	\section{Thought Experiment and Physical Dimensions}
	\label{sec:exp}
	%---------------------------
    For the sake of completeness, we show how a real physical Aqua-PACMANN XNOR could be engineered. Let us suppose that we use a squared bucket of dimension $d=10$~cm, with a water height at rest $h_0=1$~cm. In Fig.~\ref{fig:proc}, the length $d$ of our bucket is divided in $100$ points, setting our spatial unit as $D=1$~mm.     
    Using as a detector in $x_D$ a standard CCD camera, we can consider a frame rate order of magnitude as $10^3$~fps, which means that our temporal unit can be as small as $T=10^{-3}$~s. This way, each processing in Fig.~\ref{fig:proc} is completed in $0.1$~s.
    
    Equations~(\ref{eq:normalization}) are now fully determined, being $v_0=\frac{D}{T}=1$~m/s. Indeed, in shallow water, surface wave speed can be estimated as $\sqrt{gh}$, with $g$ the gravitational acceleration~\cite{2015Boyd}. In our case, $\sqrt{gh}\sim 3$~m/s, comparable to our speed unit $v_0$. The water height evolution in time can be determined as $h(x,t)=u(x,t)\times10T$, to keep the identity $r_2\times10T=h_0$.
    Following this normalization, the soliton defined in Eqs.~(\ref{eq:soliton},\ref{eq:solparameters}) has amplitude $(r_1-r_2)\times10T=1$~cm, wavenumber $k=\frac{1}{2D}=5$~cm$^{-1}$, wavelength $\lambda=4\pi D=12.6$~mm, and velocity $v=1.33$~m/s. On the other hand, the encoding waves in Table~\ref{tab:enc} have non-zero amplitudes $\epsilon_{1,2}\times10T=2.5$~mm, wavenumbers $k_1=\frac{\sqrt{3}}{4D}=4.33$~cm$^{-1}$ and $k_2=\frac{1}{2D}=5$~cm$^{-1}$, wavelengths $\lambda_1=4\pi D=14.5$~mm and $\lambda_2=4\pi D=12.6$~mm, and velocities $v_1=0.92$~m/s $v_2=0.83$~m/s. The delay between the encoding waves and the soliton is $1.28\times10^{-2}$~s and length of the excitation is $lD=2$~cm.

    %The system described until now is rather ideal. A real-world implementation of this reservoir computer should deal with other phenomena, here neglected. To start, the spatial one-dimensional approximation holds true only in large buckets, a limitation that we will overcome in our future work. Moreover, the KdV equation does not consider any wave relaxation effect, describing a system undergoing never-ending wave reflection, a phenomenon that can be neglected only in the infinite boundary approximation. However, a real hydrodynamic system, with dissipation and weak turbulence, owns such a relaxation effect. Propagation of waves in shallow fluids with dispersion, dissipation, and weak turbulence is modeled by the KBK equation~\cite{2019Alimirzaluo}. This model has several advantages that will be presented in a future paper. By the use of the KBK equation, Aqua-PACMANN RC architecture can be extended to the recurrent case. Indeed, the wave reflection timescale can be tuned by tuning the losses, an effect that can introduce a non-null matrix $\mathbf{W}$ in Eq.~(\ref{eq:RCx}), and let us move from an ELM to a proper ESN.
	
	\section{Conclusions}
    We showed the design of the Aqua-Photonic-Advantaged Computing Machine by Artificial Neural Networks~(Aqua-PACMANN), the first hydrodynamic neuromorphic computer based on nonlinear wave interaction.
    In our system, wave propagation in shallow water and their collision with a KdV soliton allow the realization of an XNOR logic gate.    
    From an algorithm perspective, the current version of Aqua-PACMANN is based on extreme learning machine.
    In the last part of this manuscript, we also trace the way to a real-world implementation. %where the reservoir computing architecture is generalized to echo state networks.
    
	\section{Conflict of Interest}
	The Authors are employed by Apoha Ltd., a company dedicated to the development of biophotonic computational devices.

	\bibliographystyle{unsrt} % We choose the "plain" reference style
	\bibliography{refs} % Entries are in the refs.bib file
	
\end{document}